%% file: main.tex
\documentclass{article}  % Comment this line out if you need a4paper
\usepackage[margin=1.0in]{geometry}
\usepackage{amsmath}
\usepackage{amssymb}
\usepackage{subcaption}
\usepackage{graphicx}
\usepackage{url}
\usepackage{color}
\usepackage{csquotes}
\usepackage{float}

\newcommand{\ie}{\textit{i.e.}, }
\newcommand{\eg}{\textit{e.g.}, }
\newcommand{\etal}{\textit{et al.}}
\newcommand{\starnote}[1]{}
\newcommand{\laststarnote}[1]{}
\newcommand{\donestarnote}[1]{}

\title{\LARGE \bf
Diver Interest via Pointing:\\ Human-Directed Object Inspection for AUVs*
}

\author{Chelsey Edge$^{1}$ and Junaed Sattar$^{2}$% <-this % stops a space
\thanks{The authors are with the Department of Computer Science and Engineering (CSE), Minnesota Robotics Institute (MnRI), University of Minnesota Twin Cities, Minneapolis, MN, USA.  
{\small\{$^{1}${\tt edge0037},$^{2}${\tt junaed}\}{\tt @umn.edu}}}%
\thanks{\textbf*This work was supported by the US National Science Foundation Award IIS-\#1637875, and the MnRI Seed Grant.}}% <-this % stops a space

\begin{document}

\maketitle
\thispagestyle{empty}
\pagestyle{empty}

%%%%%%%%%%%%%%%%%%%%%%%%%%%%%%%%%%%%%%%%%%%%%%%%%%%%%%%%%%%%%%%%%%%%%%%%%%%%%%%%
\begin{abstract}
In this paper, we present the Diver Interest via Pointing (DIP) algorithm, a highly modular method for conveying a diver's area of interest to an autonomous underwater vehicle (AUV) using pointing gestures for underwater human-robot collaborative tasks. 
DIP uses a single monocular camera and exploits human body pose, even with complete dive gear, to extract underwater human pointing gesture poses and their directions.
By extracting 2D scene geometry based on the human body pose and density of salient feature points along the direction of pointing, using a low-level feature detector, the DIP algorithm is able to locate objects of interest as indicated by the diver. 
% By extracting 2D scene geometry based on the human body pose, the DIP algorithm is able to locate an area of interest as indicated by the diver. Through the density of salient feature points along the direction of pointing using the SIFT algorithm or 
DIP makes it possible for scuba divers and swimmers to use directional cues, through pointing, to an AUV for inspection, surveillance, manipulation, and navigation. 
We examine the elements that make up our method, provide quantitative and qualitative evaluation, and demonstrate AUV actuation based on diver pointing gestures in closed-water human-robot collaborative experiments. 
Our evaluations demonstrate the high efficacy of the DIP algorithm in correctly identifying the direction of a pointing gesture and locating an object within that region of interest. We also show that the findings of the algorithm qualitatively conform with human assessment of pointing gestures, directions, and targets.

\end{abstract}

%%%%%%%%%%%%%%%%%%%%%%%%%%%%%%%%%%%%%%%%%%%%%%%%%%%%%%%%%%%%%%%%%%%%%%%%%%%%%%%%
\input{TextFiles/Introduction}

\input{TextFiles/Background}

\input{TextFiles/Method}

\input{TextFiles/Evaluation}

\input{TextFiles/Conclusion}
%\clearpage
\bibliographystyle{IEEEtran}
\bibliography{bibliography,sattar_j,allbibs}

\end{document}

%% file: TextFiles/Introduction.tex
\section{INTRODUCTION}
\begin{figure}
% \vspace{1mm}
%\vspace{3mm}
    \centering
        \includegraphics[width=.95\textwidth]{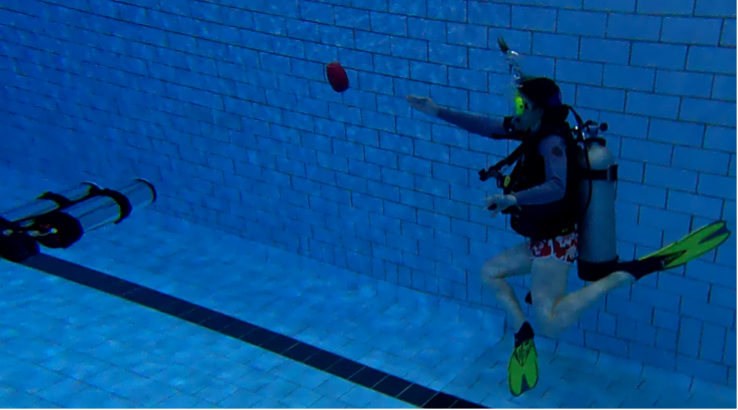}%
        \caption{LoCO AUV running DIP to navigate towards an object in response to diver pointing gesture. }%
    \label{fig:working_point}
    \vspace{-4mm} % Adjust as you see fit. 
\end{figure}

The use of autonomous underwater vehicles (AUVs) to perform tasks underwater has become increasingly relevant in recent years. 
From inspection and maintenance of underwater infrastructure~\cite{petillot_pipelone}, biological monitoring~\cite{ModasshirRobio2018_coralident}, and archaeology~\cite{bingham_archaeology}, the utility of these AUVs heavily relies on working and cooperating with human divers. 
While Remotely Operated Vehicles (ROVs) are able receive instruction through a tether, AUVs require different communication methods, such as fiducial markers~\cite{Sattar2007CRV} or gesture-based languages (\cite{islam_dynamic, Sattar2018JFR-Islam-MotionGestures, Chavez_hri}). 
\starnote{CE: cite help please, you added a lot of citations and I'm not familiar enough with them.}\laststarnote{JS: These are fine, but did you need something more specific? CE: I don't think so?}
For each of the above-mentioned scenarios, it is highly likely that a robot will be directed to specifically navigate along a particular direction or inspect an object; this is referred to as the site acquisition and scene re-inspection (SASR) task. 
Such tasks could include inspecting a specified region of a pipeline, taking pictures of particular coral for a conservation biologist, or recovering an artifact.
As many divers in these situations have their own areas of expertise and may not be robotics experts, a communication vector that is easily understood and capable of being performed naturally is essential. % to the future of this domain. 
For this reason, we introduce a system for divers that uses pointing gestures to instigate a response from an AUV towards an object or location of interest (Fig.~\ref{fig:working_point}).

Pointing is used as a natural form of human communication to share location and interest. 
Expanding this communication vector to robotics is not itself an unexplored area of interest (see Sec.~\ref{sec:background}). %\cite{Delmerico_survey2.4}. 
% Directing drones to a specified object~\cite{Medeiros2021_fire} and a wheeled companion robot to a parking space~\cite{richarz_companion_loc} through pointing are two specific examples of recent work incorporating pointing gesture understanding to human-robot interaction (HRI) and collaborative tasks. 
In the subtopic of marine robotics, however, the use and even exploration of this topic is almost non-existent (other than very few exceptions, \eg\cite{Andrea_thesis}).
Different methods of indicating interest are often used by scuba divers, \eg shining lights, carrying physical objects such as sticks or poles, and dropping strobes or beacons.
However, these methods come with significant limitations. 
The underwater domain is known for its high degree of sensory signal attenuation and dispersion, and thus methods such as shining lights on objects of interest may not be precise enough for many needs. 
Without necessitating extra tools, pointing gesture communication relies on some form of visual perception. 
Underwater vision is challenging due to natural environmental factors (\eg scattering, absorption, and refraction of light among others).
% While vision itself in most circumstances is usually available for on-land and in-air robotic tasks, underwater is known to be challenging visually due to natural environmental factors (\eg scattering, absorption, and refraction of light among others). 
% Some characteristics of the environment that distorts vision as we see it above water include distortions caused by light absorption, refraction, and attenuation. 
Image enhancement techniques (\eg\cite{Sattar2020RSS-Islam-SESR,Sattar2022RSS-Islam-SVAM}) have been shown to mitigate some of these factors and improve the ability to use RGB cameras. 
These factors still significantly impact the \emph {reliability} of commonly used terrestrial sensors, such as light-wave based Time-of-Flight RGB-D cameras, used in many pointing gesture based systems (\cite{Grossmann_Object,Shukla_ispointing,Tolgyessy_location,droeschel_TOF,abidi_loc}). 
As vision itself is challenging even in the best of water conditions, adaptation of ideas and implementation of terrestrially designed algorithms can be difficult. 
%For this reason our approach is feasible with a single RGB or greyscale camera. 

Also unique to the underwater environment is the high probability that the diver, robot, and even object of interest may have a free-floating ability. 
For example, during terrestrial and pick-and-place applications, it can generally be assumed that anything static on a ground or table environment will remain there until moved deliberately. 
Even with relation to drones, the human is potentially able to remain on the ground plane and calculations for direction can take advantage of this. 
In open water, however, it is highly unlikely that the diver will remain on the floor of the body of water; in fact, they may even be moving along with water currents. 
Our approach takes this into account by leveraging only the pose of the diver and robot view space to find the area of interest.

In this paper, we introduce \textbf{DIP} (\textit{Diver Interest via Pointing}), a method to inform an AUV of the location of an unknown object via a pointing gesture. 
We show that the use of a human pose estimator in conjunction with a single camera is sufficient to create an area of interest from which an AUV can detect an object of interest. 
In addition to locating the area of interest as indicated by a diver in still images, we provide a working example of how our method can be integrated into an object inspection system, utilizing the LoCO AUV~\cite{Sattar2020IROS-Edge-LoCO}. 

% We make the following contributions:

%  \\

% \begin{itemize}
%     \item Propose a communication vector from a diver to an AUV. Object in the frame
%     \item Implement the proposed communication vector in a modifiable way
%     \item Provide evidence of a working system on the LoCO AUV
% \end{itemize}

%  \\
%  \\
%  \\
% AUV persform tasks need to communicate/respond to human cooperators one gesture used between people is using pointing to denote a subject of interest. Robots can use theis too. have seen with drones and terrestrial vehicles

%% file: TextFiles/Background.tex
\section{Related Work}
\label{sec:background}
\donestarnote{JS: Please do not edit this section anymore. I had gone through this and now having to do this again. If you must, please ask me first?}

In this section, as the use of pointing control with AUVs is a novel idea, we focus our attention on the use of pointing gestures within the scope of robotics in general. DIP requires the use of human pose estimation, and also fits within the broader definition of human-robot interaction (HRI); as such, we also provide some context on the current state of gestural communication with AUVs.

\subsection{Gestural Communication with AUVs}
Using gestures to interface with AUVs, while relatively under-explored in comparison with aerial and terrestrial vehicles, is beneficial for divers as it requires no extra equipment. Currently deployed systems (\cite{islam_dynamic, Sattar2018JFR-Islam-MotionGestures, Chavez_hri}) have been designed with  ease of use of divers in mind. 
Expanding this interface to convey locational cues can be seen as a next step. 
While there is no current literature on directing an AUV to a location or object of interest through pointing gestures, Walker~\cite{Andrea_thesis} provides an analysis of the ability to recognize pointing gestures underwater on state-of-the-art deep-learned object detectors (\eg Single Shot Multibox Detector (SSD)~\cite{liu2016ssd}, Faster R-CNN~\cite{NIPS2015_5638}). 
% We incorporate their work into our sample system of a use case for our method. 
We contribute the first look at using pointing to provide specific locational cues of a diver's area of interest to an AUV, and exploit these cues for the purpose of AUV-assisted task completion in that location. 
\starnote{JS: I'm a little concerned with how many times you use the word `implement' which may push the reviewers to underestimate the paper's contributions.} \starnote{JS: What is this next phrase for?}\donestarnote{CE:deleted, unnecessary} %is pointing or not no vector

\subsection{Applications of Human Pose Estimation}
\begin{figure}
%\vspace{2mm}
     \centering
     \includegraphics[width=.95\linewidth]{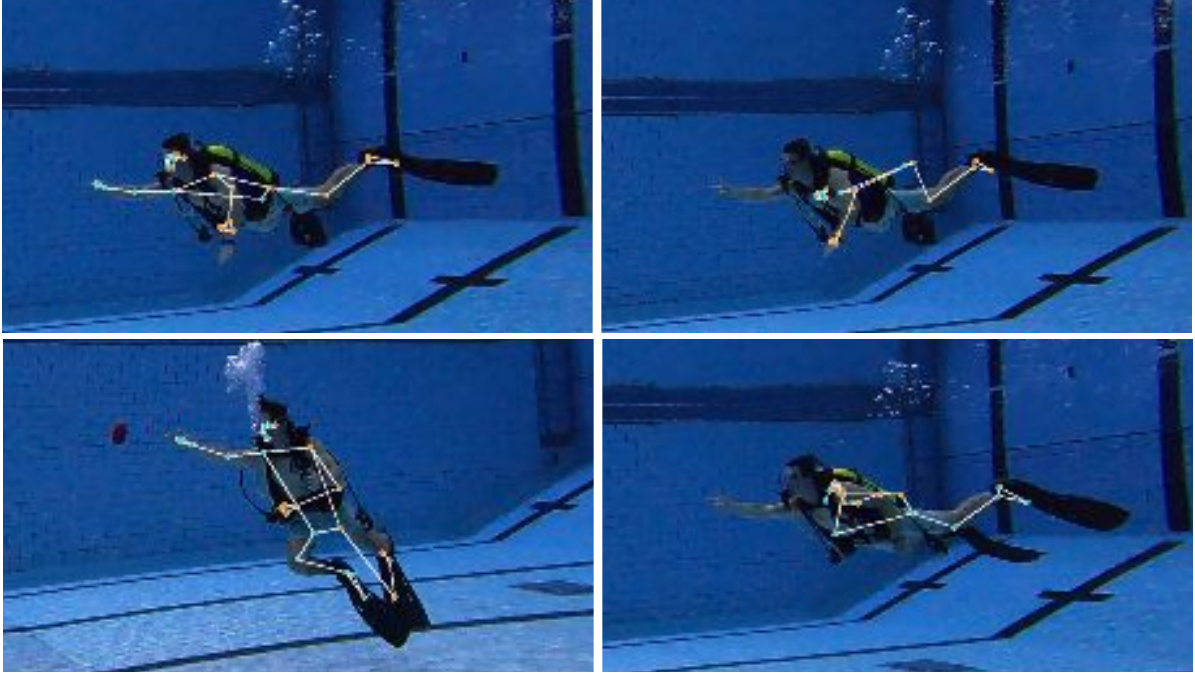}
         \caption{Example of some failure cases in underwater human pose estimation. Even in closed-water conditions, a combination of environmental factors, non-standard body pose, and SCUBA attire can affect robust `terrestrial' vision algorithms.}
        \label{fig:pose_estimator}
     \vspace{-6mm}
\end{figure}

Human pose estimators are often used to provide landmarks for pointing gestures (\cite{Tolgyessy_location, richarz_companion_loc,abidi_loc,Medeiros2021_fire}). 
The pose estimator locates landmarks, typically joints, on a human body. 
While human pose estimation in both 2D and 3D is widely studied (see~\cite{CHEN2020Survey,Andriluka_2014_survey,SARAFIANOS2016_survey}), there does not exist a vision-based human pose estimator dedicated for underwater use. 
Chavez \etal~\cite{chavez_pose} tracks a diver using point clouds of diver pose; however, specific pose landmarks are not found. 
Terrestrial-based, out-of-the-box monocular 2D pose estimators have been applied with some success; 
\eg Islam \etal~\cite{Islam2021_pose} use estimated human body pose (using OpenPose~\cite{openpose}) to find relative positions of two robots, and Fulton \etal~\cite{fulton_adroc} create an autonomous diver approach method using TRT pose~\cite{trtpose}. 
The success of human pose estimators is severely impacted not only by underwater vision degradation, but also by the positions of the human body with relation to training data and additional gear worn by scuba divers (Fig.~\ref{fig:pose_estimator}). 
We choose to incorporate the Mediapipe~\cite{mediapipe} Pose~\cite{Bazarevsky_blazepose} framework as the foundation of DIP, because of\donestarnote{Sorry for being a grammar nazi.} the cited low latency and impressive results on physical activities like yoga and dance, which also deal heavily with uncommon human poses.

\subsection{Pointing for Purpose}
In terrestrial and aerial robotics, the use pointing gestures to relay information has been widely studied. 
For example, terrestrial and aerial vehicles can be given direction to park in a certain location determined by a pointing pose estimate \cite{Tolgyessy_location, richarz_companion_loc,abidi_loc}. 
For these applications, the user is able to point at a location on the ground plane and direct the robot there. This is an impossibility in many underwater robotics applications as touching the ground plane, if it exists in view, goes against the task that is being accomplished (\eg coral reef inspection, or invasive species detection). 
Aerial and terrestrial robotics have also seen application of finding an object of interest through pointing gestures. 
Pick-and-place tasks with robotic arms have seen success through hand gesture pointing~\cite{Shukla_ispointing,Littmann_Drees_Ritter_1996}. 
Gro{\ss}mann \etal~\cite{Grossmann_Object} use a pointing gesture to indicate an object on a shelf that should be of interest to a mobile manipulator.
Medeiros \etal~\cite{Medeiros2021_fire} is able to direct a drone towards an object based on intersection of a pointing direction of the user's arm and an object's bounding box. 
Delmerico \etal~\cite{Delmerico_survey2.4} provide a survey of many more uses for pointing gestures with an emphasis on the uses with rescue robotics.
However, no work exists for the use of pointing gestures for control or communicating user interest in the underwater environment.

%% file: TextFiles/Method.tex
\section{Diver Interest via Pointing}
\begin{figure}[ht]
%\vspace{2mm}
    \centering
        \includegraphics[width=.55\linewidth]{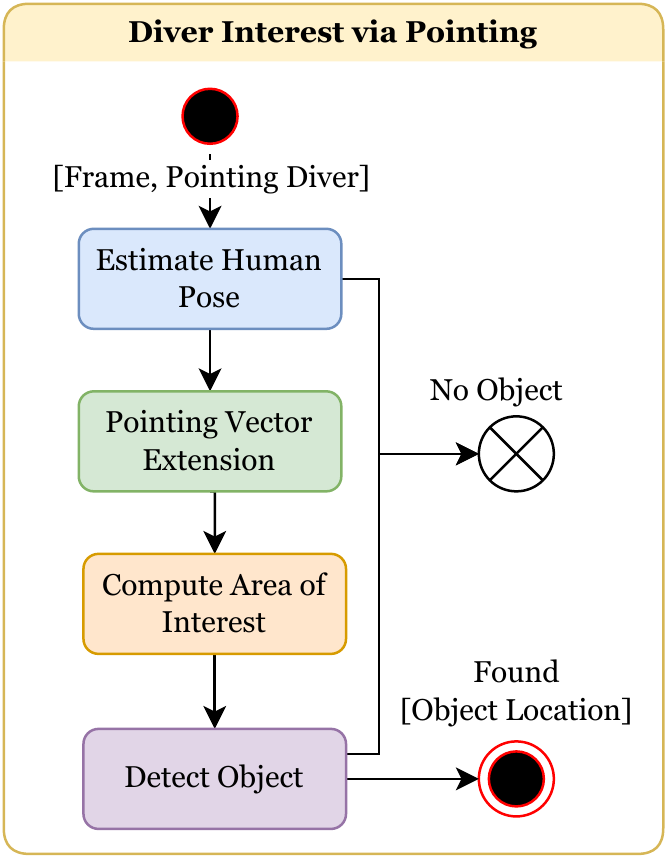}%
        \vspace{-1mm}
        \caption{Schematic diagram of the DIP algorithm. Colors denoting each step are consistent with those in Figs.~\ref{fig:pose_object_eval} and~\ref{fig:sample_system}.}%
    \label{fig:DIP algorithm}
    \vspace{-5mm} % Adjust as you see fit. 
\end{figure}

%The communication method itself presents as a heavily modifiable system. 
\laststarnote{JS:Can we just get figure 3 on this page and keep figure 4 on the next one? CE: Keep this note here until we get figures placed correctly? They keep moving around on me.}
The DIP approach is comprised of a human pose estimator, a method for the AUV to predict a diver's area of interest, and an object detection algorithm applied to the predicted area of interest (Fig.~\ref{fig:DIP algorithm}). 
In the following section we describe the method to detect a diver's pointing pose, and using that, detail the creation of the area of interest. 
In addition, we enumerate the design choices that allow the implementation of the proposed algorithm on-board a physical robot. 

For the purposes of this section we make the following assumptions:
\begin{itemize}
    \item The diver is situated in an upright, pointing pose.
    \item The diver is pointing with their right arm.%, not based on hand or finger location.
\end{itemize}
% Determining if a person underwater is pointing is outside of the scope of this project. However, as shown in Sec.~\ref{Experimental Implementation}, integration with~\cite{Andrea_thesis} is implemented to present a functioning system on-board the LoCO-AUV~\cite{Sattar2020IROS-Edge-LoCO}. All hyper-parameters are given in accordance with our experimental setups and use an image size of \textbf{480x640}.

\subsection{Diver Pose Estimation}
\donestarnote{JS: Two potential things to address; 1. how hard was it to detect those pose keypoints for divers in scuba gear, and would training a deep model help improve the accuracy? and 2. why are we not doing 3D vector pointing in this case?}

AUV communication using pointing gestures from divers requires robust estimation of human pose. Human pose estimation underwater is challenging for a variety of reasons including naturally degraded vision, non-standard body positions and the wearing of SCUBA gear by divers (see Fig.~\ref{fig:pose_estimator}). 
To minimize the need for correctly positioned body landmarks, we use only two, the elbow and wrist.  
As the connection between these two points must be linear (\ie anatomically, there are no joints between them in human physiology), this creates the ability to generalize potential locations of interest in a rather straightforward manner. 
By simply extending the line segment connecting the elbow to the wrist, the general pointing direction can be found.
The shoulder is excluded as, unless the diver's body is turned towards the camera, landmark detection can be difficult (Fig.~\ref{fig:pose_estimator}). 
We explicitly avoid the use of the hands or fingers for finding the pointing direction, as the poor visibility underwater, particularly at the usual interaction distance between an AUV and a diver, can make it difficult for pose estimators designed for terrestrial use to identify those body parts.
For the purpose of our project, we use the Mediapipe Pose~\cite{mediapipe} estimator, although any pose estimator which can provide wrist \(w_{(x,y)}\) and elbow \(e_{(x,y)}\) landmarks in the two-dimensional image space can be used. 
Once the pose with the wrist and elbow landmarks have been located, we are able to determine a direction and local area of interest. 

The use of 2D pose for determining pointing direction for SASR type tasks is sufficient, as objects of interest will often be close to the diver in distance and we can approximate the 3D directional vector in 2D image space. As marine environments are also generally uncluttered and without many structures, the need for depth information about the scene is not necessary for many tasks. Particularly when coupled with a robust object detector, 2D pointing vectors will often suffice for such purposes. Future work looks at expanding DIP to include 3D information for use in specialized situations.

\subsection{Determining the Area of Interest}
\label{sec:aoi}
\begin{figure}[ht]
    \centering
        \includegraphics[width=.75\linewidth]{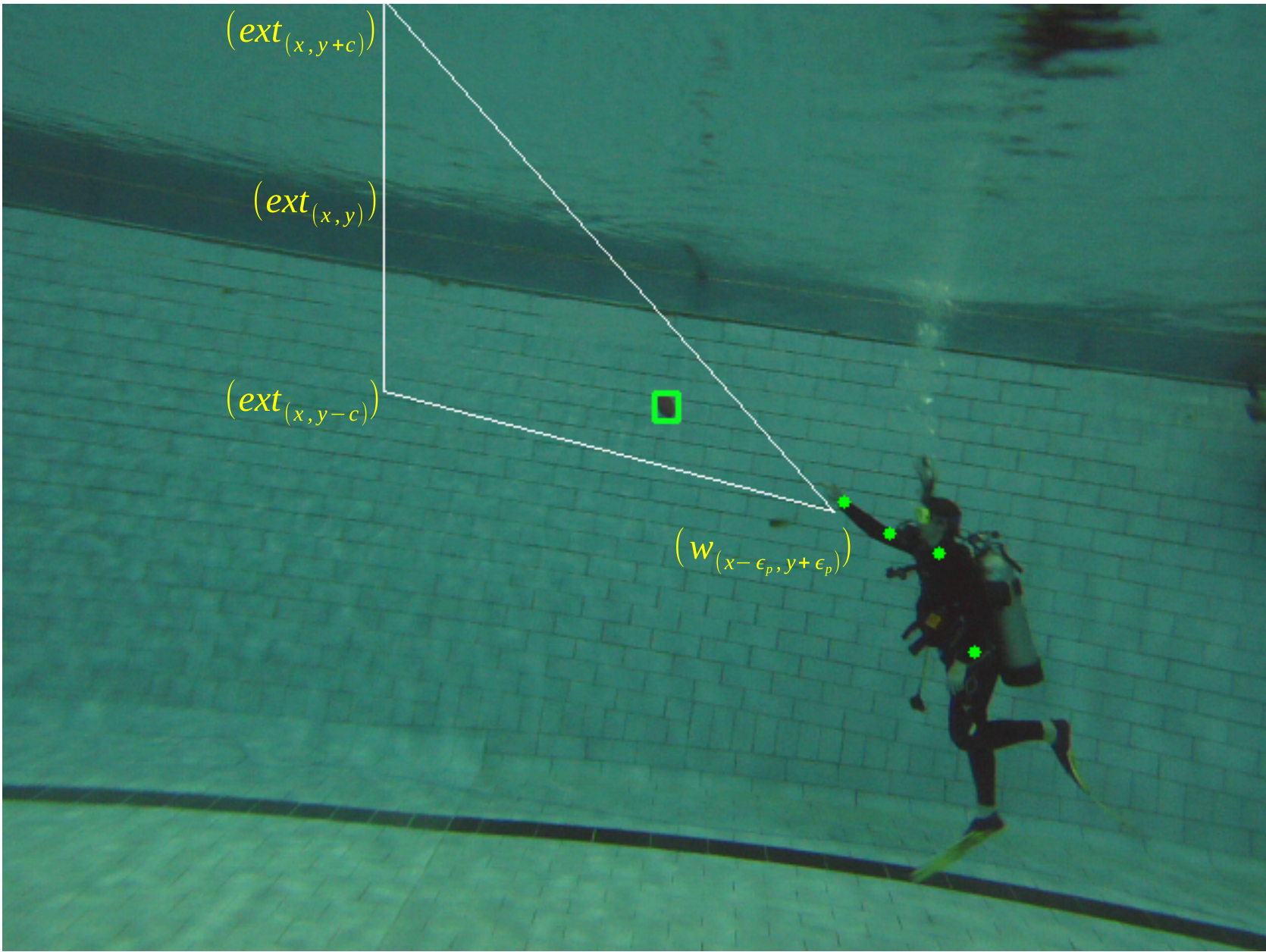}%
        \vspace{-1mm}
        \caption{Visualization of the DIP algorithm: Human pose landmarks are located, an area of interest is created based on pose pointing vector, and an object is located within the area of interest.}%
    \label{fig:DIP_success}
    \vspace{-5mm} % Adjust as you see fit. 
\end{figure}

The area of interest to the robot is defined to be a triangular region extending from the diver's wrist, shown in Fig.~\ref{fig:DIP_success} as the white triangle. 
A triangular shape is chosen, as the farther away a diver is from an object, the less accurate the actual gesture may be (\ie the diver's own gesture pointing at the object of interest will be less accurate from farther away). 
This error can potentially be magnified in relation to above-water scenarios, as the object, diver, and robot itself may not be able to remain stationary for a variety of reasons, such as being suspended in the water, moving with a current, or constantly maneuvering to hold position. %due to swimming and water movements. 
Choosing a triangular region to search makes it possible to account for such inherent inaccuracies in pointing and have a more accurate detection of the object of interest. 

% The definition of this  location of this triangle is dependent on the direction and length of the line segment from the diver's elbow to their wrist using landmarks from the pose estimator. 

First, the line segment connecting the elbow and wrist landmarks (as defined by the human pose estimator) is extended \(ext_{(x,y)}\) by a scaling factor \(sf\) which can be modified as needed (Eq.~\ref{eq:point_extension}): 

\begin{equation} \label{eq:point_extension}
\centering
  \begin{aligned}
    ext_{(x)} = w_{(x)} + sf * (w_{(x)}-e_{(x)}) \\
    ext_{(y)} = w_{(y)} + sf * (w_{(y)}-e_{(y)}),  
  \end{aligned}
\end{equation}
A scale factor of $10$ has empirically been found to be sufficient in our investigations.

Once the segment is extended, the resulting point \(ext_{(x,y)}\) is defined as the end of the pointing vector. 
To create an area of interest in the shape of a triangle in image space, two vertices are defined by adding and subtracting a \textit{vertical constant} $c$, to the $y$ value of the point extension (\eg$ext_{(x,y \pm c)}$) (see Fig.~\ref{fig:DIP algorithm} for a visualization). 
% midpoint of the base of the triangular area of interest.
% a triangle with a height of the segment from \(w_{(x,y)}\) to \(ext_{(x,y)}\). 
% The length of the base can, again, be modified as necessary based on image size. 
% We use a base approximately equal to half of the image height all experiments. 
% With this constant, we take into account that the further away an object is, the greater the potential for human pointing error in an unstable, moving environment. 

The third vertex of the triangle is defined by the wrist landmark of the human pose detector. This vertex may also be offset by a small amount $\epsilon_p$ to reduce false positives of hand detection to compensate for detector errors.
%With a trained detector, the offset may be omitted. 
The vertices of the triangle are therefore defined as: $(w_{(x-\epsilon_p,y+\epsilon_p)}),(ext_{(x,y \pm c)})$. 
% The white triangle in Fig.~\ref{fig:DIP_success} visualizes the diver's area of interest. 
% % 
% A constant of $100$ has empirically been found to be sufficient for $c$, although this varies based on image and object size.For our experiments, we offset the wrist vertex by 5 pixels in order to reduce false positives of detecting the hand. 
% % With a trained detector, the offset may be omitted. 
% The vertices of the triangle are therefore \(w_{(x-5,y+5)}\)\(ext_{(x,y \pm 100)}\)See \textbf{figure} for an example of an area of interest. 

Once the area of interest is computed, search for the object of interest via object detection is confined to this area. 
% If the type of the object is known \emph{a prioiri}, a pre-trained object detector can be used to find it. 
% FOLLOW UP 
%  location. %From this point, any object detection method can be used.

\subsection{Detecting the Object of Interest}
\label{subsection:detection method}
As the main goal is to identify the diver's intent to highlight an area of interest, the consequent object detection step is highly dependant on the actual task given to the AUV and thus should be changed accordingly, \ie a trained trash detector should be used if the AUV's task is to locate trash in an area of interest.
However, even without knowing the object type ahead of time, it is possible to exploit low-level image features to identify objects pointed at by divers within the triangular area of interest. 
For example, point features (\eg SIFT~\cite{Lowe04SIFT}, ORB~\cite{rublee2011orb}) and edge detectors (\eg Canny~\cite{canny1986computational}) can be used to identify regions with high probability of being objects of interest. 
The motivation to use such low-level feature extractors stems from the fact that the underwater environment often has little background variation, and only salient features will be on or in close proximity to the object. 
Due to the challenges of detection in an underwater environment, we demonstrate detection through two methods: keypoint and contour detection. Via keypoint detection, we choose the \emph{keypoint with the greatest strength} to represent the object, even if this has the potential for occasional to false positive detections. 
The Canny edge detector is used to extract object contours. If the centroid of the contoured region is within the triangle of interest, the target is chosen as the object of interest. 
Fig.~\ref{fig:DIP_success} shows an example of DIP successfully finding the object pointed to by the diver.

%% file: TextFiles/Evaluation.tex
\section{Evaluation}
\label{sec:evaluation}

Obtaining ground truth is difficult in underwater environments as there is constant motion between diver, robot, and objects. 
In addition, accuracy of DIP as a whole rather than the accuracy of the pose and object point location is a necessity\starnote{Do you mean to say it's more important that DIP works better overall than it's components individually?CE:Yes, don't know if that makes sense}\donestarnote{Yep, works, just doing a minor edit.}; therefore, for the following evaluations, visual cues rather than straightforward landmark matching will be used in assessment. 

All evaluations are performed using the same parameters with input image sizes of $640\times 480$ pixels. \laststarnote{JS: I missed this last pass -- what hyperparameters? aren't these for deep nets?CE: Yes, I wasn't sure what to call them. They are the maybe just parameters? Confusion }
We use an empirically found vertical constant of $100$. 
In other words, the height of the triangular area interest will be $200$ pixels, just under half the image height.\laststarnote{What is this vertical constant? Is this the height of the triangle?: Yes} We also offset the wrist vertex by $5$ pixels (\ie $\epsilon_p=5$, see Sec.~\ref{sec:aoi}) to reduce false positives of detecting the hand. 
The vertices of the triangular area of interest therefore become $(w_{(x-5,y+5)}),(ext_{(x,y \pm 100)})$.

\subsection{DIP and Human-based Pointing Correlation}\label{subsection:Human Pointing}
\begin{figure}
\vspace{1mm}
     \centering
     \includegraphics[width=.98\linewidth]{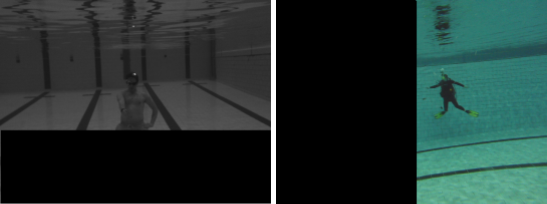}
 
        \caption{Two of the eight images the human correlation study participants were asked to annotate. The locations of the target objects are blackened out so as to not bias the participants' assessment of the pointing direction.}
        \label{fig:evalseen_img}
\vspace{-2mm} 
\end{figure}

\begin{figure}
     \centering
     \vspace{2mm}
        \includegraphics[width=.95\linewidth]{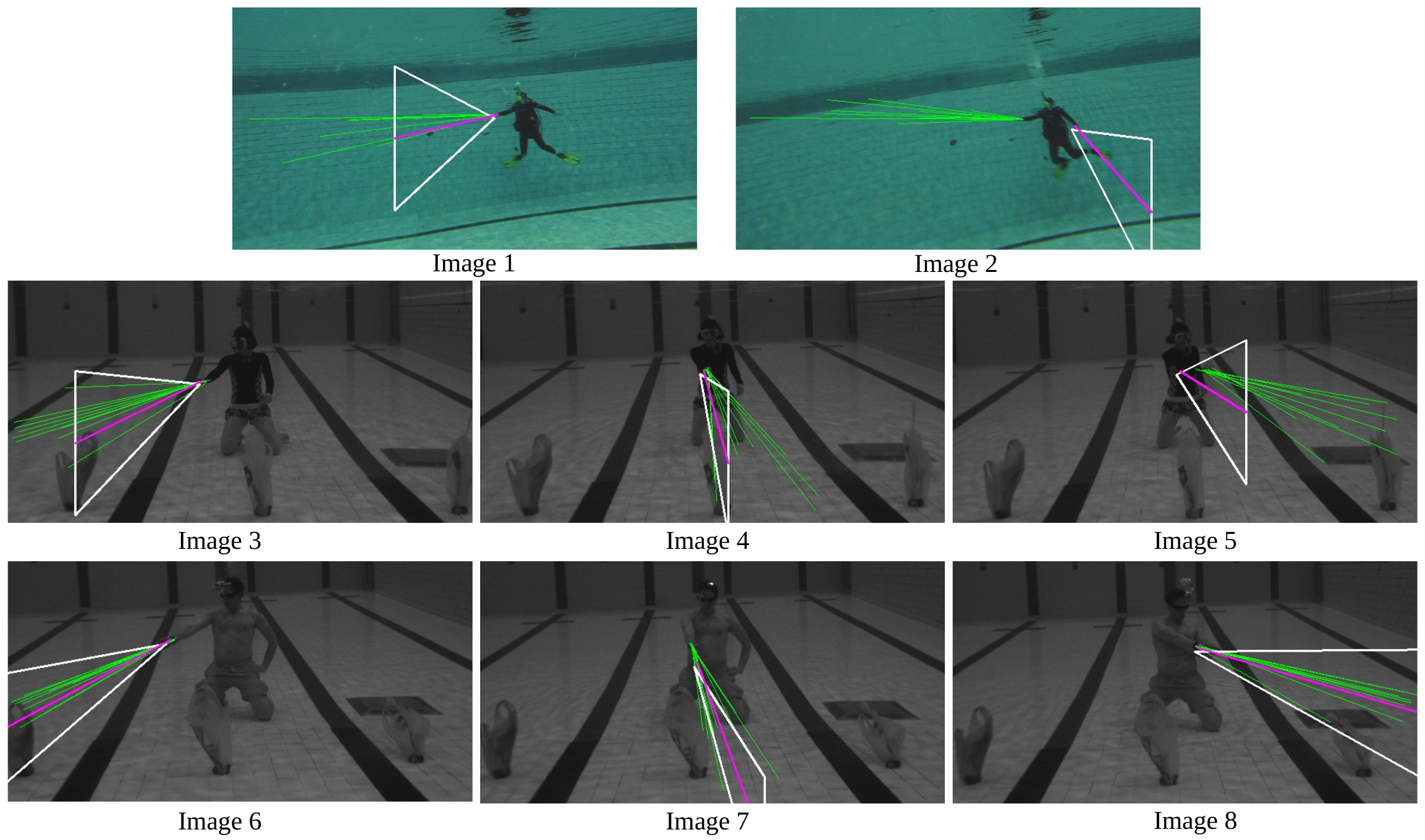}  
        \vspace{-2mm}
        \caption{Results of human-annotated vectors (green) shown with the pose estimator-based resultant vector (pink), and DIP area of interest (white triangle). DIP's output predominantly aligns with human assessment of pointing directions.}
    \label{fig:eval_studyresults}    
\end{figure}

For the first experiment, we present a small human study in which participants annotate a pointing vector on $8$ images. 
Each image consists of a submerged diver pointing towards an object, either suspended in the water, or resting on the bottom.
The general location of an object of interest was removed (Fig.~\ref{fig:evalseen_img}) in order to prevent preconceptions of what the diver is pointing towards. 
Nine participants annotated a line segment of the assumed pointing vector, beginning at the diver's wrist. 
Fig.~\ref{fig:eval_studyresults} shows a compilation of the results: green line segments represent the human-annotated vectors, the pink segment represents DIP's foundational pointing vector, and the white triangle represents the location of diver interest as produced by DIP. 
As can be seen from the quantitative data in Table~\ref{table:point_variance}, where image numbers match the vector images,  fitting a 2D vector to a point can be highly variable even by human annotators. 
% In the table, image numbers correspond to Fig.~\ref{fig:eval_studyresults}. 
% \textit{Mean} refers to the mean angle of study participants' annotations, \textit{DIP} is the angle of the vector our area of interest is based on, \textit{Difference} is the difference between the \textit{Mean} and \textit{DIP} angle measures, and \textit{variance} is the variance of the human-annotated vectors. 
Angles are not included for DIP where pose estimation is incorrect.

Fig.~\ref{fig:eval_studyresults} also conveys some of the challenging situations even within a highly-specified underwater setting. 
As our method is dependent on human pose, if the pose captured is incorrectly, the area of interest will be so as well. 
Failure cases were found to occur frequently when the diver's arm is located in front of the trunk as seen in center and diver left-pointing evaluation cases (See Fig.~\ref{fig:eval_studyresults}). \donestarnote{refer to images in Fig~\ref{fig:eval_studyresults} please?}

\begin{table}
\centering
\vspace{2mm}
\begin{tabular}{ c|c|c|c|c } 
 
  & \shortstack{Mean \\ (Human)}  & DIP  & Difference &\shortstack{Variance \\ (Human)} \\ 
  \hline   \hline
   Image 1 & $3.059$ rad & $2.912$ rad & $ 0.147$ rad & $0.008 $   \\
   \hline
Image 2 & $3.198$ rad & --- & --- & $0.003 $   \\% $0.845$ 
\hline
 Image 3 & $2.849$ rad & $2.677$ rad & $0.121$ rad & $0.017$  \\
   \hline
 Image 4 & $ 1.097$ rad & $1.309$ rad & $0.211$ rad & $0.069$  \\ 
   \hline
Image 5 & $0.357$ rad & --- & --- & $0.024$  \\ %$0.559$
  \hline
Image 6 & $2.717$ rad & $2.643$ rad & $0.074 $ rad & $0.004$     \\
  \hline
Image 7 & $1.249$ rad & --- & --- & $0.028 $   \\
  \hline
 Image 8 & $0.286$ rad & $0.288$ rad & $0.002$ rad & $0.008$   \\
\end{tabular}
\caption{Results of human study (Fig.~\ref{fig:eval_studyresults}): Mean human-annotated angle measure, DIP angle measure (when pose is correct), Difference in Mean and DIP angles, and variance in human-annotated responses are recorded. All angles are in radians.}
\label{table:point_variance}
\vspace{-5mm} 
\end{table}

\subsection{DIP-based Pose and Object Detection}

\donestarnote{Are we talking about the actual implementation details anywhere? You mention SIFT at the end of the section below but I don't recall you clearly stating what the implementation was based on.: This is done in Sec.~\ref{subsection:detection method}, added re-emphasis below }

We evaluate DIP for the ability to use human pose estimation and low-level feature detectors to first compute a diver's specified region of interest and then locate an unknown object within that region.
% In this evaluation, we visually inspect
~\donestarnote{Yes but you are reporting hard numbers here. I think we should rephrase this `visually inspect' wording somewhat.} %the results of a closed water test of DIP.
Due to the challenging environment, we consider the ability to identify an accurate region of interest more beneficial than exact landmark matching. \donestarnote{Elaborate on what you mean by `funtionality' a bit, please?}
In regards to unknown object detection, we provide results for both the SIFT~\cite{Lowe04SIFT} algorithm and Canny~\cite{canny1986computational} edge detection as defined in Sec.~\ref{subsection:detection method}. 

\begin{figure}[t]
\vspace{2mm}
    \centering
    \begin{subfigure}[t]{0.24\textwidth}
        \centering
        \includegraphics[width=\textwidth]{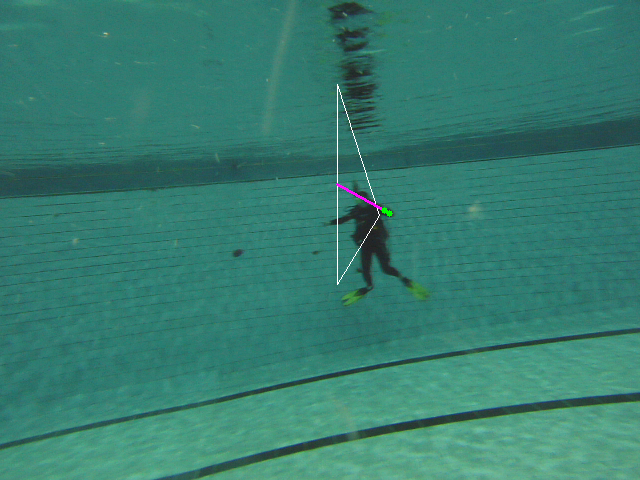}
        \caption{Incorrect human pose found.}
    \end{subfigure}
    \hfill
    \begin{subfigure}[t]{0.24\textwidth}
        \centering
        \includegraphics[width=\textwidth]{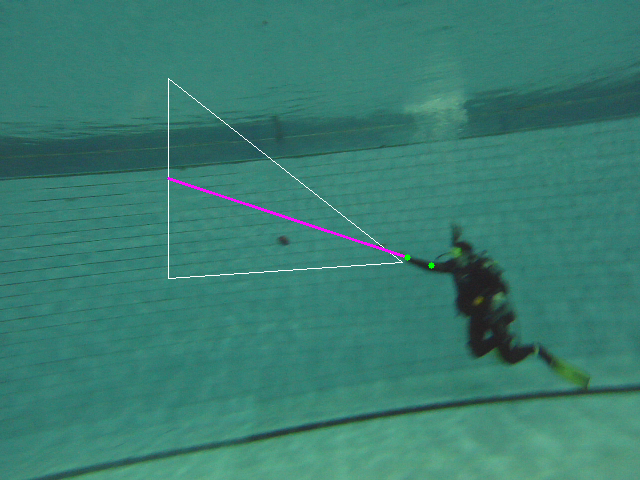}
        \caption{DIP success.}
    \end{subfigure}
    \hfill
    \begin{subfigure}[t]{0.24\textwidth}
        \centering
        \includegraphics[width=\textwidth]{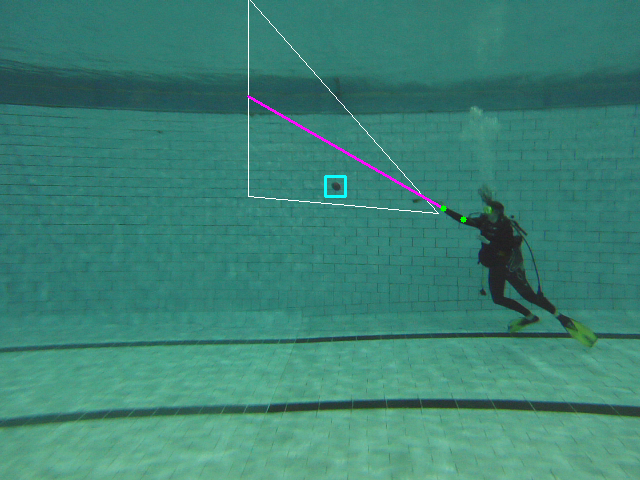}
        \caption{Unknown object located during DIP through the SIFT algorithm.}
    \end{subfigure}
    \hfill
    \begin{subfigure}[t]{0.24\textwidth}
        \centering
        \includegraphics[width=\textwidth]{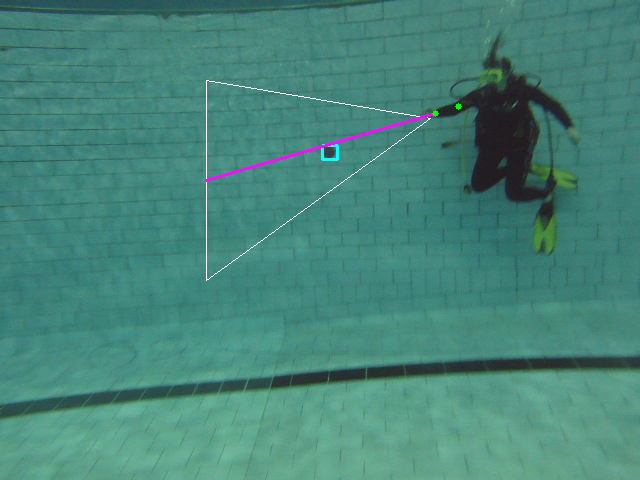}
        \caption{Unknown object located during DIP using Canny edge detection.}
    \end{subfigure}
    \caption{Samples from the pose and object detection evaluation image set. All images are sourced from the LoCO AUV and may not be considered good quality to human viewers.}
\label{fig:loco_examples}
% \vspace{-7mm}
\end{figure}

\begin{figure}[ht]
\vspace{-1mm}
    \centering
        \includegraphics[width=.75\linewidth]{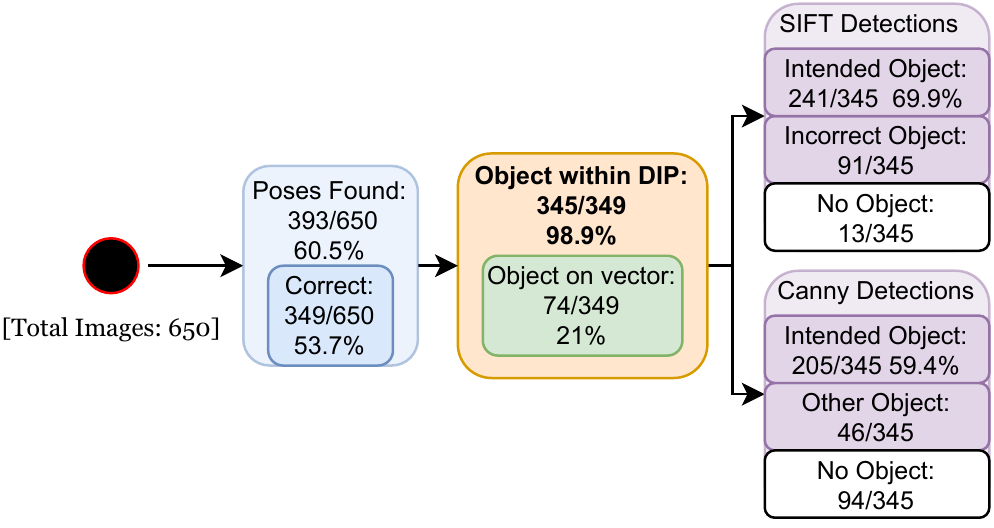}
        \vspace{-2mm}
        \caption{Evaluation of DIP pose estimation and object detection. The ratios presented depend on success of the previous steps. Colors denote the step evaluated as in Fig.~\ref{fig:DIP algorithm}.}%
    \label{fig:pose_object_eval}
    \vspace{-2mm}
\end{figure}

A selection of $650$ frames were taken from a ROS~\cite{quigley_ros_2009} bag file recorded with the LoCO AUV~\cite{Sattar2020IROS-Edge-LoCO} in a closed-water environment (Fig.~\ref{fig:loco_examples}). 
All images in sequence are included unless the entirety of the diver's body is not in the scene. 
Fig.~\ref{fig:pose_object_eval} shows results of a visual examination of this dataset. 
While not all images may be considered \enquote{good quality} by human standards, we include these images to demonstrate the effectiveness of DIP from the AUV viewpoint. 
The evaluation dataset can be considered to represent fairly optimal conditions in terms of water quality and distance from the diver. 

When taking into account only those images that produce a \enquote{correct} pose ($349$/$650$ images), we see that the object of interest is included in the region of interest $345$ out of $349$ times or \textbf{$98.85$\%}.  
On the other hand, the pointing vector itself lands on the object only $74$ times, or \textbf{$21.2$\%}. 
As supported by the results of Sec.~\ref{subsection:Human Pointing},\donestarnote{which previous section?} due to variations in pointing, locating an area of interest produces better detection results than choosing a single vector. 
With regards to the evaluation of object detection, the inclusion of a good object detector is essential for specific tasks. 
However, due to the nature of the underwater environment, objects of interest are able to be located far better than random chance with low-level features, with the SIFT algorithm failing to locate a potential object of interest in only $13$ out of $345$ images.~\laststarnote{SIFT just finds 13? The phrasing is confusing. What are we trying to tell the people here? Are those 13 detection the erroneous ones?: Correct, but reworded so hopefully makes more sense.}

% Intent with this project is to show the value of the area of interest, not to test the feasibility of unknown object detectors in a pool environment. These detection results are given as a guideline to show detection in the region is possible. \textbf{This should be kept in mind when reading this section}. Also of note,  For example a wrist landmark may be considered close enough even if it is closer to the hand.   

% \begin{center}
% \begin{tabular}{ |c|c| } 
%  \hline
%  Total Images & 650  \\
%  \textbf{Pose Found} & 393  \\ 
%  Pose Correct & 349  \\ 
%  Object exists on pointing vector & 74   \\
%  Object exists on or within area bounds & 345   \\
%  Canny Edge Detector & 207  \\
%  SIFT Features & 242  \\
%  \hline
% \end{tabular}
% \label{table:pose_eval}
% \end{center}

\subsection{Runtime}
\donestarnote{General comments: Please enclose numbers from results in math mode \$ \$ delimeters. And, make sure there's a space before you begin a parenthesis.}

With Mediapipe Pose as a backbone, the DIP algorithm (pose estimation and SASR creation) runs at $4.714$ fps on an Intel\textsuperscript{TM} i7-6700 CPU, which is acceptable for AUV operations. 
%As DIP is used to compute a location of interest and commence object detection, not itself needed to maintain tracking, it runs sufficiently quickly.
\laststarnote{Another distraction case? I do not understand this sentence.: Do I even need justification? JS: Justification as to why it can run fast? IMHO, no. CE:In that case it gets removed. }

\section{Experimental Implementation onboard the LoCO AUV}
\label{sec:experimental_implementation}

\donestarnote{Okay, this is specific to the robot implementation, correct?:Yes}

\begin{figure}
 \vspace{-2mm}
    \centering
        \includegraphics[width=.8\linewidth]{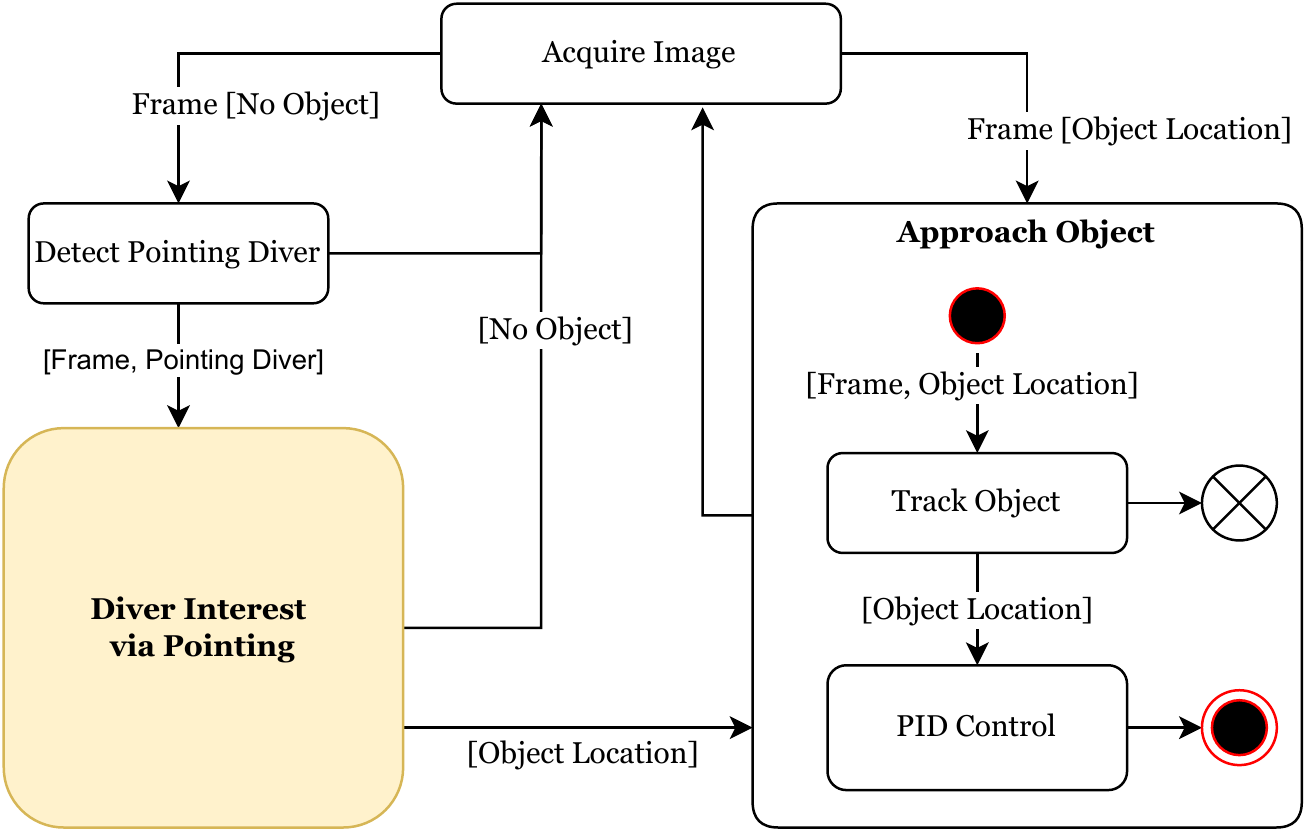}%
        \vspace{-1mm}
        \caption{Diagram showing the full system of Unknown Object Investigation guided by the DIP algorithm, as implemented on-board the LoCO AUV.}%
    \label{fig:sample_system}
    \vspace{-5mm}
\end{figure}

DIP works as the guiding force for an unknown object investigation task (Fig.~\ref{fig:sample_system}). 
We deploy DIP, along with the rest of the system on the LoCO AUV. 
The system runs entirely onboard the LoCO AUV, uses a monocular RGB camera, and performs computations on an NVIDIA Jetson TX2 embedded system.
In an enclosed pool environment, LoCO was required to detect that a diver is pointing, employ DIP to determine where an object may be located and locate an object, then finally actuate towards the object for closer inspection. 

% Hyperparameters for this experiment are set for input image sizes of $640\times 480$ pixels. 
% We use a vertical constant of $100$. \starnote{What is this vertical constant?}
% We also offset the wrist vertex by $5$ pixels (\ie $\epsilon_p=5$, see Sec.~\ref{sec:aoi}) to reduce false positives of detecting the hand. 
% The vertices of the area of interest therefore becomes: $(w_{(x-5,y+5)}),(ext_{(x,y + 100)}),(ext_{(x,y - 100)})$.

\begin{figure}
\vspace{-5mm}
     \centering
     \begin{subfigure}[t]{0.48\textwidth}
         \centering
         \includegraphics[width=\textwidth]{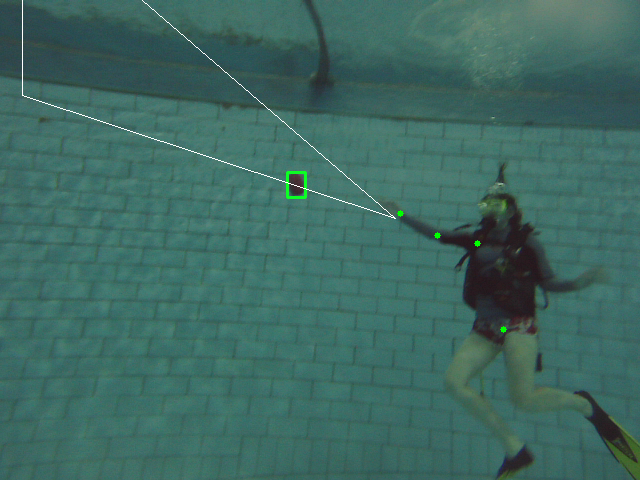}
         \caption{An object is identified through the DIP algorithm.}
         \label{feg:detected}
     \end{subfigure}
     \hfill
     \begin{subfigure}[t]{0.48\textwidth}
         \centering
         \includegraphics[width=\textwidth]{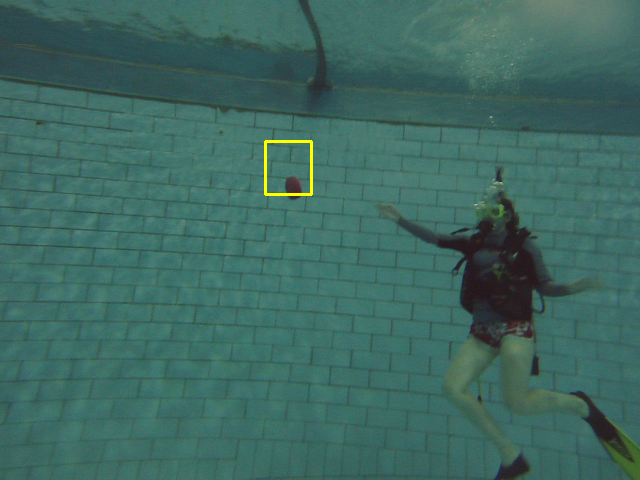}
         \caption{The object is tracked in the following frame.}
         \label{fig:tracked}
     \end{subfigure}
        \caption{Demonstration of the task: Unknown Object Investigation. Images are from the LoCO AUV point of view.}
        \label{fig:rostopic_figure}
        \vspace{-7mm}
\end{figure}

Within this system, a diver is determined to be pointing through a pre-trained SSD~\cite{liu2016ssd} detector with a VGG-16~\cite{Simonyan15VGG16} backbone~\cite{Andrea_thesis}. 
Once the pointing diver is detected, DIP proceeds to check incoming frames until the pose is detected.  \laststarnote{why is `DIP' in bold? CE: Good question, now it's not.}
In the same frame the pose is detected, object detection is attempted. 
This cycle proceeds on successive input frames until an object is located. 
At this point, the diver's interest and the object within the location is considered confirmed and the algorithm moves to initiate robot locomotion. 
The AUV begins moving towards the object through a Proportional-Integral-Derivative (PID) approach controller defined through a bounding box and image size ratio. 
We use the CAMShift tracker~\cite{opencv_library} to return a new bounding box location for the PID controller to continue AUV movement towards the object as it might be subjected to unintended motion underwater. 
Fig.~\ref{fig:rostopic_figure} shows snapshots from LoCO's viewpoint during a successful trial. 
% Faster tracking algorithms and PID adjustments will be highly dependent on future AUV tasks.
\donestarnote{Potentially remove this sentence.:Have done so}
Fig.~\ref{fig:sample_system} diagrams the system as a whole as it runs onboard the LoCO AUV.
% shows the design of the system as it is  of the system.

Out of five trial runs, the area of interest is defined (\eg the human pose is correct) four times. 
An object was located each of these times and LoCO proceeded to move autonomously towards the perceived object, making task execution successful. 
Three out of four times, however, markings on the pool wall were located inside the area of interest. These markings were considered to be the object and LoCO moved in that direction. \laststarnote{3 out of 4? Ah I see, LoCO moved to the `wrong' object, then? Gotcha, one last thing: the markings were in the correct area of interest, right? If so, we should say that in the sentence. CE:Done} 
Using a detector trained for specific objects of interest could help mitigate this issue.

%% file: TextFiles/Conclusion.tex
\section{Conclusion}
We present a novel communication algorithm, Diver Interest via Pointing (DIP), to signal directional intent to an AUV. By detecting the natural communication vector of pointing, we show that an AUV is able to infer a diver's area of interest. While working in the challenging underwater environment, we have shown that a triangular area of interest can be found with the use of a monocular camera and out-of-the-box human pose estimator. Knowing that the area of interest exists, an AUV is able to perform tasks specifically within that area as determined by a diver. One such task is the inspection of an unknown object. We validate DIP through an integrated system and demonstrate its feasibility onboard the LoCO AUV. Future work will continue to improve directional gesture communication through the use of 3D scene geometry, pointing gesture classification for downstream robot tasks such as manipulation and data collection, and design of robot-to-human feedback to reduce ambiguity in determining objects of interest. 

%%as well as answering the question \enquote{Now that an AUV can compute an area of interest, what will it be possible for the AUV do with new information?}